\documentclass[conference]{IEEEtran}
\usepackage{array}
\usepackage{graphicx}
\usepackage[english]{babel}
\usepackage{pgfplots}

\usepackage{tikz}

\usetikzlibrary{arrows,shapes,positioning,shadows,trees}

\tikzset{
  basic/.style  = {draw, text width=10cm, drop shadow, font=\sffamily, rectangle},
  root/.style   = {basic, rounded corners=2pt, thin, align=center,
                   fill=white!30},
  level 2/.style = {basic, rounded corners=6pt, align=center, fill=white!60,
                   text width=9.4em},
  level 3/.style = {basic, thin, align=left, fill=white!60, text width=6.5em}
}

\begin{document}

\title{Virtual Machine Placement Literature Review}
\author{\IEEEauthorblockN{Fabio L\'opez Pires}
\IEEEauthorblockA{Itaipu Technological Park (PTI)\\National University of Asunci\'on (UNA)\\fabio.lopez@pti.org.py\\Paraguay}
\and
\IEEEauthorblockN{Benjam\'in Bar\'an}
\IEEEauthorblockA{National University of Asunci\'on (UNA)\\East National University (UNE)\\bbaran@pol.una.py\\Paraguay}
}

\maketitle

\hyphenation{re-le-vant con-so-li-da-tion know-led-ge cons-traints mea-su-ra-ble pro-blems fo-re-cas-ting con-si-de-red con-si-de-ring}

\begin{abstract}
Cloud Computing Datacenters host millions of virtual machines (VMs) on real world scenarios. In this context, Virtual Machine Placement (VMP) is one of the most challenging problems in cloud infrastructure management, considering also the large number of possible optimization criteria and different formulations that could be studied. VMP literature include relevant topics such as energy-efficiency, Service Level Agreements (SLA), cloud service markets, Quality of Service (QoS) and carbon dioxide emissions, all of them with high economical and ecological impact. This work presents an extensive up-to-date review of the most relevant VMP literature in order to identify research opportunities.
\end{abstract}
\IEEEpeerreviewmaketitle

\section{Background and Motivation}
% Open indentation until last sections
{\setlength{\parindent}{12pt}	
The process of selecting which virtual machines (VMs) should be located (i.e. executed) at each physical machine (PM) of a datacenter is known as Virtual Machine Placement (VMP).

The VMP problem has been extensively studied in cloud computing literature and several surveys have already been presented. Existing surveys focus on specific issues such as: (1) energy-efficient techniques applied to the problem \cite{beloglazov2012energy,salimian2013survey}, (2) particular architectures where the VMP problem is applied, specifically federated clouds \cite{gahlawat2014survey} and (3) methods for comparing performance of placement algorithms in large on-demand clouds \cite{mills2011comparing}.

Beloglazov et al. presented in \cite{beloglazov2012energy} a survey of energy-aware resource allocation policies and scheduling algorithms considering QoS. The following open challenges were identified considering energy-aware management of cloud computing datacenters: (1) development of fast energy-efficient algorithms for the VMP, considering multiple resources for large-scale systems with the ability to predict workload peaks to prevent performance degradation, (2) energy-aware optimization of virtual network topologies between VMs for optimal placement in order to reduce network traffic and thus energy consumed by the network infrastructure, (3) development of new thermal management algorithms to appropriately control temperature and energy consumption, (4) development of workload-aware resource allocation algorithms, considering that current approaches assume a uniform workload, and (5) decentralization and distributed approaches to provide scalability and fault-tolerance to the VMP problem resolution.

Salimian et al. presented a review of different selection and placement algorithms for energy-efficient management of cloud computing datacenters \cite{salimian2013survey}. Approaches for virtual and physical resources modeling, applied techniques and future work were identified for each studied article. Most relevant future work include: (1) investigation of VMP for multi-core architectures considering multiple resources, (2) consideration of dynamic thresholds for QoS and (3) development of intelligent schemes according to workload and considering live migration.

Gahlawat et al. proposed in \cite{gahlawat2014survey} a brief survey of the main cloud federation architectures and approaches considered for VMP problem formulation in this particular scenario. It is important to remember that cloud federation is the practice of voluntarily interconnecting cloud infrastructures of different Cloud Service Providers (CSPs), mostly to respond to workload peaks.

Finally, Mills et al. presented in \cite{mills2011comparing} an objective method to compare performance of placement algorithms in large on-demand clouds, where 18 different algorithms for VMP were compared considering 39 variables such as: reallocation rate, user request rate, allocation rate and disk space utilization, just for cite a few.

The above mentioned surveys and research articles focused into specific issues related to the VMP problem. To the best of the author's knowledge, there is not existing research work that presents a general and extensive study of a large part of the VMP literature. This work presents an extensive up-to-date review of the most relevant VMP literature in order to identify research opportunities on this important and promising research area. 

\section{Reviewed Literature}
A selection process of current research work was defined and performed in order to study a large part of the most relevant VMP literature for this survey. This section presents a detailed description of the literature selection process which is summarized in Figure \ref{figure_literature}.

\subsection{Keywords Search}
The selection process of relevant articles started with a search of research articles from Google Scholar database [scholar.google.com] with at least one of the following selected keywords in the article title:
(1) virtual machine placement, (2) vm placement, (3) virtual machine consolidation, (4) vm consolidation or (5) server consolidation. 

This keywords search step results in 446 research articles. A detailed list of the 446 resulting articles could be found in \cite{lopez2014survey_data}, specifically in the worksheet named keywords search.

\subsection{Publisher Filtering}
Considering the large number of results from keywords search step, the literature selection process focused in research articles published in the following relevant publishers: (1) ACM, (2) IEEE, (3) Elsevier and (4) Springer. Percentage of articles per publisher in the studied universe is summarized in Figure \ref{figure_publisher}.

This publisher filtering step results in a reduction from 446 to 172 (38.6\%) research articles of the literature. A detailed list of the 172 resulting articles could be found in \cite{lopez2014survey_data}, specifically in the worksheet named publisher filtering.

\subsection{Abstract Reading}
Considering the 172 resulting articles from the publisher filtering step, an abstract reading was performed in order to identify only the most relevant articles that specifically study the VMP problem. After the abstract reading, 93 research articles were selected from the VMP literature. A detailed list of the 93 resulting articles could be found in \cite{lopez2014survey_data}, specifically in the worksheet named abstract reading. 

Finally, short papers (i.e. research articles with less than 6 pages) were removed from the selected literature, resulting in 84 selected articles of the VMP literature for the detailed study presented in this survey.

\begin{figure}[!b]
\centering
\includegraphics[width=0.5\textwidth]{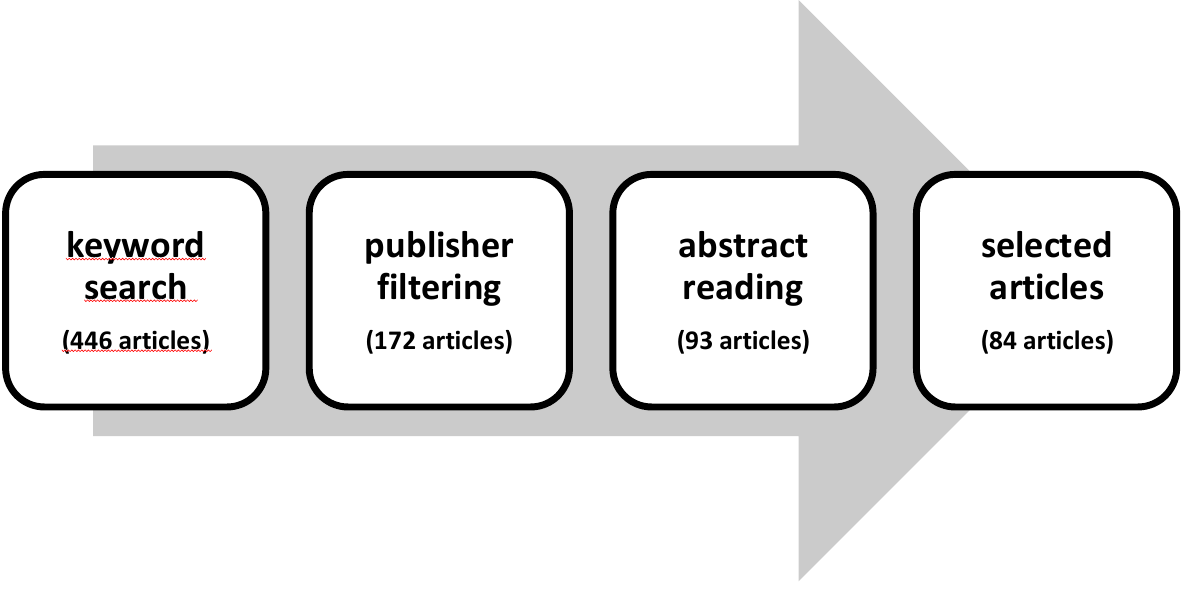}
\caption{VMP literature selection process}
\label{figure_literature}
\end{figure}

\begin{figure}[!t]
\centering
\newcommand{\slice}[4]{
  \pgfmathparse{0.5*#1+0.5*#2}
  \let\midangle\pgfmathresult

  % slice
  \draw[thick,fill=black!10] (0,0) -- (#1:1) arc (#1:#2:1) -- cycle;

  % outer label
  \node[label=\midangle:#4] at (\midangle:1) {};

  % inner label
  \pgfmathparse{min((#2-#1-10)/110*(-0.3),0)}
  \let\temp\pgfmathresult
  \pgfmathparse{max(\temp,-0.5) + 0.8}
  \let\innerpos\pgfmathresult
  \node at (\midangle:\innerpos) {#3};
}

\begin{tikzpicture}[scale=2]
\newcounter{a}
\newcounter{b}
\foreach \p/\t in {11/ACM, 69/IEEE, 15/Springer, 5/Elsevier}
  {
    \setcounter{a}{\value{b}}
    \addtocounter{b}{\p}
    \slice{\thea/100*360}
          {\theb/100*360}
          {\p\%}{\t}
  }
\end{tikzpicture}
\caption{Percentage of articles per publisher in the studied 84 papers.}
\label{figure_publisher}
\end{figure}
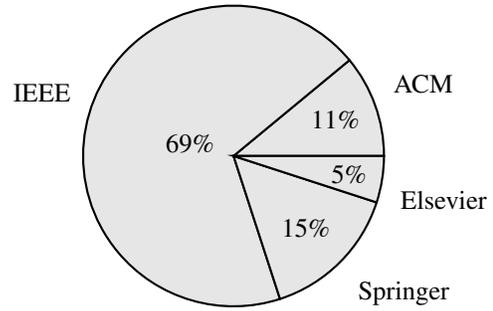

\section{Literature Overview}
Consolidating multiple underutilized servers into a fewer number of non-dedicated servers can respond to the common problem of server sprawl in datacenters \cite{Gupta2008}. The consolidation can be modeled as a variant of the bin packing problem (items to be packed are the VMs and bins are the PMs) in order to help resource management considering that until 2008, server consolidation exercise was primarily a manual process. A manual process of this nature is time consuming and depends on the subjective assessment of the decision maker. Particularly, in \cite{Gupta2008} it is studied the server consolidation problem considering technically deprecated PMs into technically superior PMs, but this problem could also be studied as a VMP problem. Experimental tests are performed considering a maximum number of 1000 PMs for mapping into 620 PMs, solving extremely large instances of the problem in a reasonable amount of time, i.e. 0.188 seconds against 6262.1 seconds (1.74 hours) for the optimal solution. The proposed algorithm obtained solutions with a 1.5 ratio more PMs than the optimal solution for largest problem instances, i.e. 620 PMs obtained against 411 PMs from optimal solution.

If a VM is considered to be a three dimensional object (CPU, memory and storage) then the problem of placing the VMs over the PMs looks similar to the three dimensional bin packing problem, but they are not exactly the same. In 3D bin packing problem, a set of three dimensional objects (generally cuboids) are required to be placed inside three dimensional containers (also cuboids). The aim is to pack as many objects in the containers as possible, so that the number of containers required is minimized. While packing objects into a given container, two objects can be placed side by side or one on top of the other, but if we consider VMs as the objects then placing VMs side by side or one on top of the other is not a valid operation. This is because once a resource is utilized or occupied by a VM, it can not be reused by any other VM \cite{Mishra2011}. The VMP problem is actually similar to Vector Packing Problem which is also a NP-Hard problem.

In \cite{Xu2010}, VMs and PMs are represented as a $d$-dimensional vector where each dimension corresponds to one type of resource (e.g. CPU, memory and storage) and for each PM should be balanced along different dimensions. Operational power and thermal dissipation are also studied. For each objective an efficiency metric was defined. For the resolution of the VMP formulation, authors proposed a Genetic Algorithm (GA). Genetic operators are: (1) selection and crossover, where a random process do not perform efficiently, so the selection and crossover is based on each of the three efficiency metrics resulting on better solutions, (2) mutation, where groups of VMs are randomly selected and eliminated.

In \cite{chaisiri2009optimal} are studied two payment plans in cloud computing markets: (1) reservation and (2) on-demand plans (Amazon EC2 and GoGrid). Considering that prices of resources in reservation plans are generally cheaper than in on-demand plans, in advance reservation should be applied as maximum as possible. On-demand plans only should be used when the reservation does not fully meet predicted workload. The proposed algorithm makes a decision to host a certain number of VMs on appropriate cloud providers obtaining an optimal solution from a stochastic integer programming (SIP) formulation considering two stages of decision making: (1) first stage defines the number of VMs provisioned in reservation phase, while (2) the second stage defines the number of VMs allocated in both utilization and on-demand phases.

In \cite{Bin2011}, it is defined a new high-availability property for a VM; when a VM is marked as $k$-resilient, it should be guaranteed that it can be relocated to a non-failed host without relocating other VMs as long as there are up to $k$ PMs failures. Anti-location, anti-co-location and resources constraints are also proposed for the VMP formulation, considering the optimization of the problem in its dual space (dual optimization problem). The transformed formulation enable a CP engine to compute a solution in reasonable time on practical cases.

When a client request comes, the infrastructure service provider needs to identify a PM to place the client VM so that the following objectives can be achieved: (1) the SLA associated with the VM is satisfied, and (2) the performance metrics of SLAs associated with the VMs previously running on the target physical node is least affected \cite{Do2011}. The proposed prediction contains the following initiation steps: (1) collect the performance data of a given application during its previous executions and treat it as foreground performance metrics, (2) collect the performance data of candidate physical nodes and treat the data as background performance metrics, and (3) perform Canonical Correlation Analysis (CCA) on the performance data collected in the previous two steps and produce the canonical weight vectors and scores for the most correlated patterns.

A formal model for the application (to better understand structural constraints) and the (hierarchical) datacenter (to effectively capture capabilities) is proposed in \cite{Jayasinghe2011}. Demand, communication and availability constraints are considered for the VMP formulation, considering experiments for a small datacenter (60 PMs) and a large one (200 PMs).

Services offered by cloud providers such as Amazon, Microsoft, IBM, and Google are implemented on thousands of servers spread across multiple geographically distributed datacenters. There are at least three reasons behind this geographical distribution: the need for high availability and disaster tolerance, the sheer size of the computational infrastructure, and the desire to provide uniform access times to the infrastructure from widely distributed user sites \cite{Le2011}.

In \cite{Tsakalozos2011}, it is implemented a two-phase optimization process. During the first phase a subset of PMs with properties that best serve the VM placement is selected. The goal in selecting a subset of all available PMs is to reduce the number of constraints and the search space during the second phase. In \cite{Dupont2012}, it is proposed a framework for: (1) being able to add or modify a constraint without changing the algorithms and (2) being able to test and activate a new algorithm without having to re-implement every constraint within it.

\section{Optimization Approaches}\label{optimization_approaches}
This section presents the optimization alternatives proposed in the studied articles according to the optimization approach of the studied objective functions. The identified optimization approaches may be classified as: (1) mono-objective optimization problem (MOP), (2) multi-objective solved as mono-objective (MAM) and (3) pure multi-objective (PMO). The mentioned optimization approaches are detailed in the following sub-sections and summarized in Table \ref{table_optimization_approaches}.

\subsection{Mono-Objective Approach}
A mono-objective approach considers the optimization of only one objective or the individual optimization of more than one objective function, one at a time.

According to the studied articles, the research of the VMP problem has been mainly guided by the mono-objective optimization approach considering that 61.9\% of the studied articles proposed a mono-objective approach (MOP) for solving the VMP problem. From the 52 articles that studied the VMP problem in a mono-objective approach, almost 40 different objective functions were proposed. It is also remarkable that some works studied the same objective functions but proposed different modeling approaches (e.g. economical revenue maximization could be achieved by minimizing the total economical penalties for SLA violations \cite{dang2013higher}, by minimizing operational costs \cite{Huang2012b,Huang2012} or even by maximizing the total profit for leasing resources \cite{Shi2011}). 

Considering the large number of proposed objective functions and different approaches for modeling an objective function, multi-objective optimization \cite{coello2002evolutionary} could result in better and more realistic formulations of the VMP problem, optimizing more than just one objective function at a time (e.g. achieve economical revenue maximization by simultaneously minimize the total economical penalties for SLA violations, minimize operational costs and maximize the profit for leasing resources).

\subsection{Multi-Objective solved as Mono-Objective Approach}
In this work, the optimization of multiple objective functions combined into one objective function is considered a multi-objective approach solved as mono-objective (MAM). A disadvantage of this approach is that requires a deep knowledge of the problem domain to allow a correct combination of the objective functions, which in most cases is not possible \cite{baran2005multi}.

According to this survey, 34.5\% of the studied articles proposed a multi-objective approach but finally solved the VMP as a mono-objective optimization problem. In the last years, a growing number of articles have proposed formulations of the VMP problem with this hybrid approach (see Figure \ref{figure_optimization_approach_year}).

The most popular method for solving a MAM problem is the Weighted Sum Method \cite{wang2014eqvmp,adamuthe2013multiobjective,Dong2013,fang2013power,anand2013virtual,dong2013energy,dong2013virtual,fang2013vmplanner,Shigeta2013,chen2013intelligent,prevost2013optimal,hong2013qoe,dalvandi2013time,Wang2012,Kantarci2012,Wu2012a,Jin2012,Jiang2012a,Calcavecchia2012}. In this method one objective function is formulated as a linear combination of multiple objectives.

Solving a multi-objective optimization problem as mono-objective considering the other objective functions as constraints of the problem is also a studied method in the VMP literature \cite{Sato2013,Ferreto2011a}. 

Kord et al. proposed a two-objective approach and made a trade-off between these two goals using a fuzzy Analytic Hierarchy Process (AHP) \cite{Kord2013}. Other research articles employed fuzzy logic to provide an efficient way for combining conflicting objectives and expert knowledge. For multiple objectives, fuzzy logic allows mapping of different objectives into linguistic values characterizing levels of satisfaction \cite{song2014optimization,Huang2012a,Xu2010}.

\subsection{Pure Multi-Objective Approach}
A general pure multi-objective optimization problem (PMO) includes a set of $p$ decision variables, $q$ objective functions, and $r$ constraints. Objective functions and constraints are functions of decision variables. In a PMO formulation, $x$ represents the decision vector, while $y$ represents the objective vector. The decision space is denoted by $X$ and the objective space as $Y$. These can be expressed as \cite{coello2002evolutionary}:
\\\\
\noindent \textit{Optimize:}
\begin{equation}\label{eq1} y = f (x) = [f_1(x), f_2 (x),...,f_q(x)] \end{equation}

\noindent \textit{subject to:}
\begin{equation}\label{eq2} e(x) = [e_1(x), e_2 (x),...,e_r(x)] \ge 0 \end{equation}

\noindent \textit{where:}
\begin{equation}\label{eq3} x = [x_1, x_2,..., x_p ] \in X \end{equation}
\begin{equation}\label{eq4} y = [y_1, y_2,...,y_q ] \in Y \end{equation}
\\
It is important to remark that optimizing, in a particular problem context, can mean maximizing or minimizing. The set of constrains $e(x) \ge 0$ defines the set of feasible solutions $X_f \subset X$ and its corresponding set of feasible objective vectors $Y_f \subset Y$. The feasible decision space $X_f$ is the set of all decision vectors $x$ in the decision space $X$ that satisfies the constraints $e(x)$, and it is defined as:

\begin{equation}\label{eq5} X_f = \{x \mid x \in X \wedge e(x) \ge 0\} \end{equation}

The feasible objective space $Y_f$ is the set of the objective vectors $y$ that represents the image of $X_f$ onto $Y$ and it is denoted by:

\begin{equation}\label{eq6} Y_f = \{y \mid y = f(x) \quad \forall x \in X_f\} \end{equation}

To compare two solutions in a multi-objective context, the concept of Pareto dominance is used. Given two feasible solutions $u$, $v \in X$, $u$ dominates $v$, denoted as $u \succ v$, if $f(u)$ is better or equal to $f(v)$ in every objective function and strictly better in at least one objective function. If neither $u$ dominates $v$, nor $v$ dominates $u$, $u$ and $v$ are said to be non-comparable (denoted as $u \sim v$).

A decision vector $x$ is non-dominated with respect to a set $U$, if there is no member of $U$ that dominates $x$. The set of non-dominated solutions of the whole set of feasible solutions $X_f$, is known as optimal Pareto set $P^*$. The corresponding set of objective vectors constitutes the optimal Pareto front $PF^*$.

In the considered literature, only 3.6\% of the articles proposed a multi-objective approach, using Pareto Dominance for comparing conflicting objective functions \cite{lopez2013virtual,pires2013multi,Gao2013}. To the best of our knowledge, there is no many-objective optimization proposed for the VMP problem in literature (i.e a multi-objective optimization problem with more than three objective functions \cite{von2014survey}).

\begin{table}[!t]
\renewcommand{\arraystretch}{1.3}
\centering
\caption{Optimization Approaches}
\label{table_optimization_approaches}
\begin{tabular}{|m{2.9cm}|m{1.2cm}|m{3.4cm}|}
\hline
\bfseries Optimization Approach & \bfseries Percentage & \bfseries References\\
\hline
\centering Mono-Objective (MOP) & \centering 61.9\% & \cite{tang2014hybrid,Zhang2013,Chang2013,Caron2013,li2013migration,georgiou2013exploiting,lu2013qos,alicherry2013optimizing,wang2013particle,singh2013reduce,shi2013provisioning,dang2013higher,ribas2013pbfvmc,hwang2013hierarchical,Huang2013,Jin2013,Li2013a,li2013energy,gupta2013hpc,guo2013shadow,moreno2013improved,kakadia2013network,tsai2013prevent,Zamanifar2012,Huang2012b,Wu2012b,Wu2012,Li2012a,Jiang2012,Lin2012,Biran2012,Huang2012,Dias2012,Goudarzi2012,Ribas2012,Gupta2012,Dupont2012,Tsakalozos2011,Jayasinghe2011,Le2011,Bin2011,Mishra2011,Do2011,Ho2011,Shi2011,mark2011evolutionary,Piao2010,Machida2010,Speitkamp2010,Meng2010,chaisiri2009optimal,Gupta2008}\\
\hline
\centering Multi-Objective solved as Mono-Objective (MAM) & \centering 34.5\% & \cite{wang2014eqvmp,song2014optimization,cao2014energy,zhang2014dynamic,adamuthe2013multiobjective,Sun2013,Dong2013,fang2013power,anand2013virtual,dong2013energy,dong2013virtual,fang2013vmplanner,Shigeta2013,chen2013intelligent,Kord2013,Sato2013,prevost2013optimal,hong2013qoe,dalvandi2013time,Wang2012,Kantarci2012,Wu2012a,Huang2012a,Jin2012,Jiang2012a,Calcavecchia2012,Ferreto2011,Ferreto2011a,Xu2010}\\
\hline
\centering Pure Multi-Objective (PMO) & \centering 3.6\% & \cite{Gao2013,pires2013multi,lopez2013virtual}\\
\hline
\end{tabular}
\end{table}

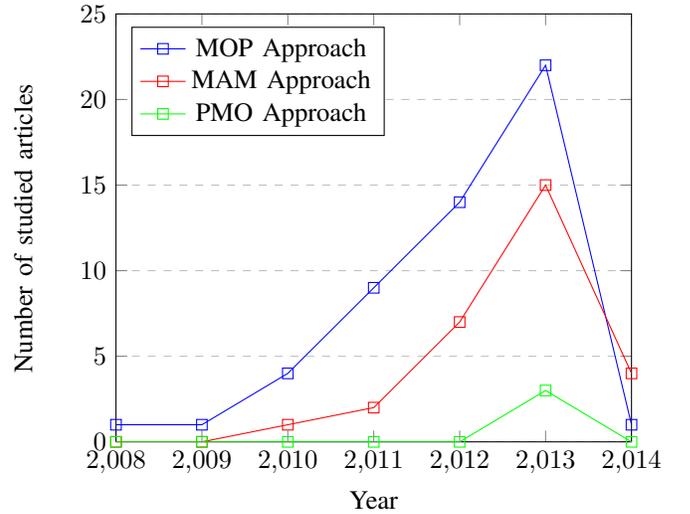
\begin{figure}[!t]
\centering
\begin{tikzpicture}
\begin{axis}[
    xlabel={Year},
    ylabel={Number of studied articles},
    xmin=2008, xmax=2014,
    ymin=0, ymax=25,
		xtick={2008,2009,2010,2011,2012,2013,2014},
    ytick={0,5,10,15,20,25},
    legend pos=north west,
    ymajorgrids=true,
    grid style=dashed,
]
 
\addplot[
    color=blue,
    mark=square,
    ]
    coordinates {
    (2008,1)(2009,1)(2010,4)(2011,9)(2012,14)(2013,22)(2014,1)
    };
		\label{MOP Approach}
		\addlegendentry{MOP Approach}

\addplot[		
		color=red,
    mark=square,
    ]
    coordinates {
    (2008,0)(2009,0)(2010,1)(2011,2)(2012,7)(2013,15)(2014,4)
    };
		\label{MAM Approach}
		\addlegendentry{MAM Approach}
		
\addplot[		
		color=green,
    mark=square,
    ]
    coordinates {
    (2008,0)(2009,0)(2010,0)(2011,0)(2012,0)(2013,3)(2014,0)
    };
    \label{PMO Approach}
		\addlegendentry{PMO Approach}

\end{axis}
\end{tikzpicture}
\caption{Articles per year according to each considered optimization approach from the universe of 84 papers.}
\label{figure_optimization_approach_year}
\end{figure}

\section{Objective Functions}\label{objective_functions}
In cloud computing datacenters with a considerable amount of PMs and VMs, there are many criteria that can be considered when selecting a possible solution for the VMP problem, depending on policies and optimization objectives. These criteria can even change from one period of time to another, which implies a variety of possible formulations of the problem and different objective functions to be optimized.

The studied articles proposed 56 different objective functions (Tables \ref{table_energy} - \ref{table_resource}) for the three optimization approaches presented in Section \ref{optimization_approaches}. Considering the large number of proposed objective functions, this work classifies objective functions with similar goals into 5 objective function groups that are presented in following sub-sections.

\subsection{Energy Consumption Minimization}\label{energy}
Considering the studied articles, 50.0\% proposed energy consumption minimization for the VMP problem considering different energy modeling approaches which are detailed below. Objective functions classified in this optimization group are: (1) Datacenter Power Consumption Minimization, (2) Energy Consumption Minimization, (3) Energy Efficiency Maximization, (4) IP Layer Power Consumption Minimization, (5) Network Power Consumption Minimization, (6) Number of PMs Minimization, (7) Power Consumption Minimization and (8) WDM layer Power Consumption (see Table \ref{table_energy}).

\begin{table}[!t]
\renewcommand{\arraystretch}{1.3}
\centering
\caption{Objective Functions - Energy Consumption Minimization Group}
\label{table_energy}
\begin{tabular}{|m{5cm}|m{2.5cm}|}
\hline
\bfseries Objective Function & \bfseries References\\
\hline
Datacenter Power Consumption Minimization & \cite{Kantarci2012}\\
\hline
Energy Consumption Minimization & \cite{tang2014hybrid,wang2014eqvmp,cao2014energy,zhang2014dynamic,fang2013power,dong2013energy,dong2013virtual,wang2013particle,singh2013reduce,Shigeta2013,Huang2013,chen2013intelligent,Sato2013,li2013energy,moreno2013improved,tsai2013prevent,Wu2012a,Huang2012a,Goudarzi2012,Dupont2012}\\
\hline
Energy Efficiency Maximization & \cite{prevost2013optimal}\\
\hline
IP Layer Power Consumption Minimization & \cite{Kantarci2012}\\
\hline
Network Power Consumption Minimization & \cite{fang2013power,dong2013energy,fang2013vmplanner,Wu2012a,Huang2012a}\\
\hline
Number of PMs Minimization & \cite{Sun2013,anand2013virtual,ribas2013pbfvmc,Wu2012b,Jiang2012,Jin2012,Ribas2012,Ferreto2011,Ho2011,Ferreto2011a,Machida2010,Gupta2008}\\
\hline
Power Consumption Minimization & \cite{pires2013multi,lopez2013virtual,Dong2013,Gao2013,fang2013power,dong2013energy,fang2013vmplanner,dalvandi2013time,Wu2012,Wu2012a,Huang2012a,Xu2010}\\
\hline
WDM layer Power Consumption & \cite{Kantarci2012}\\
\hline
\end{tabular}
\end{table}

\subsection{Network Traffic Minimization}
From the studied articles, 30.9\% proposed network traffic minimization for the VMP problem. As presented in section \ref{energy}, network traffic optimization could be studied jointly with energy consumption. Objective functions classified in this optimization group are: (1) Average Traffic Latency Minimization, (2) Cloud QoE Maximization (Response Time Minimization), (3) Cloud Service Response Time Minimization, (4) Data Access Minimization, (5) Data Transfer Time Minimization, (6) End-to-end Delay Minimization, (7) Link Congestion Minimization, (8) Migration Number Minimization, (9) Migration Overhead Minimization, (10) Migration Time Minimization, (11) Network Cost Minimization, (12) Network Performance Maximization, (13) Network Traffic Minimization, (14) Network Utilization Minimization, (15) Node Cost Minimization, (16) Overall Communication Cost Minimization, (17) WAN Communication Minimization and (18) Worst Case Cut Load Ratio Minimization (see Table \ref{table_network}).

Modeling and quantification of live migration have been studied considering the minimization of migration number as well as the minimization of the overhead and migration time (see Table \ref{table_network}). Live migration it self can be considered a very challenging future direction in the VMP context, considering it as a enabling technology for cloud computing management systems.

\begin{table}[!t]
\renewcommand{\arraystretch}{1.3}
\centering
\caption{Objective Functions - Network Traffic Minimization Group}
\label{table_network}
\begin{tabular}{|m{5cm}|m{2.5cm}|}
\hline
\bfseries Objective Function & \bfseries References\\
\hline
Average Traffic Latency Minimization & \cite{Meng2010}\\
\hline
Cloud QoE Maximization (Response Time Minimization) & \cite{hong2013qoe}\\
\hline
Cloud Service Response Time Minimization & \cite{Chang2013}\\
\hline
Data Access Minimization & \cite{alicherry2013optimizing}\\
\hline
Data Transfer Time Minimization & \cite{Zamanifar2012,Piao2010}\\
\hline
End-to-end Delay Minimization & \cite{fang2013power}\\
\hline
Link Congestion Minimization & \cite{dong2013virtual}\\
\hline
Migration Number Minimization & \cite{Calcavecchia2012,Ferreto2011a}\\
\hline
Migration Overhead Minimization & \cite{anand2013virtual}\\
\hline
Migration Time Minimization & \cite{Ferreto2011}\\
\hline
Network Cost Minimization & \cite{song2014optimization,Wang2012,Jiang2012a}\\
\hline
Network Performance Maximization & \cite{kakadia2013network}\\
\hline
Network Traffic Minimization & \cite{dong2013virtual,Shigeta2013,Zhang2013,Dias2012,wang2014eqvmp,lopez2013virtual,pires2013multi}\\
\hline
Network Utilization Minimization & \cite{georgiou2013exploiting}\\
\hline
Node Cost Minimization & \cite{Jiang2012a}\\
\hline
Overall Communication Cost Minimization & \cite{Jayasinghe2011}\\
\hline
WAN Communication Minimization & \cite{chen2013intelligent}\\
\hline
Worst Case Cut Load Ratio Minimization & \cite{Biran2012}\\
\hline
\end{tabular}
\end{table}

\subsection{Economical Costs Optimization}
From the studied articles, 22.6\% proposed economical costs optimization for the VMP problem. Objective functions classified in this optimization group are: (1) Economical Revenue Maximization, (2) Electricity Cost Minimization, (3) Operational Cost Minimization, (4) Reservation Cost Minimization, (5) Server Cost Minimization, (6) SLA Violations Minimization, (7) Thermal Dissipation Costs Minimization and (8) Total Infrastructure Cost Minimization (see Table \ref{table_economical}).

\begin{table}[!t]
\renewcommand{\arraystretch}{1.3}
\centering
\caption{Objective Functions - Economical Costs Optimization Group}
\label{table_economical}
\begin{tabular}{|m{5cm}|m{2.5cm}|}
\hline
\bfseries Objective Function & \bfseries References\\
\hline
Economical Revenue Maximization & \cite{adamuthe2013multiobjective,pires2013multi,lopez2013virtual,shi2013provisioning,hong2013qoe,Shi2011}\\
\hline
Electricity Cost Minimization & \cite{Le2011}\\
\hline
Operational Cost Minimization & \cite{Huang2012b,Huang2012}\\
\hline
Reservation Cost Minimization & \cite{chaisiri2009optimal}\\
\hline
Server Cost Minimization & \cite{Speitkamp2010}\\
\hline
SLA Violations Minimization & \cite{zhang2014dynamic,Dong2013,dang2013higher,Kord2013,prevost2013optimal,dalvandi2013time}\\
\hline
Thermal Dissipation Cost Minimization & \cite{Xu2010}\\
\hline
Total Infrastructure Cost Minimization & \cite{mark2011evolutionary}\\
\hline
\end{tabular}
\end{table}

\subsection{Performance Maximization}
In this work, objective functions that proposed formulations related to performance represent 16.7\% of the studied articles. Objective functions classified in this optimization group are: (1) Availability Maximization, (2) CPU Demand Satisfaction Maximization, (3) Deployment Plan Time Minimization, (4) Performance Maximization, (5) QoS Maximization, (6) Resource Interference Minimization, (7) Security Metrics Maximization, (8) Shared Last Level Cache (SLLC) Contention Minimization and (9) Total Job Completion Time Minimization (see Table \ref{table_performance}).

\begin{table}[!t]
\renewcommand{\arraystretch}{1.3}
\centering
\caption{Objective Functions - Performance Maximization Group}
\label{table_performance}
\begin{tabular}{|m{5cm}|m{2.5cm}|}
\hline
\bfseries Objective Function & \bfseries References\\
\hline
Availability Maximization & \cite{lu2013qos,Wang2012,Shigeta2013,Bin2011}\\
\hline
CPU Demand Satisfaction Maximization & \cite{Calcavecchia2012}\\
\hline
Deployment Plan Time Minimization & \cite{Tsakalozos2011}\\
\hline
Performance Maximization & \cite{kakadia2013network,gupta2013hpc,Gupta2012,Do2011}\\
\hline
QoS Maximization & \cite{Sato2013}\\
\hline
Resource Interference Minimization & \cite{Lin2012}\\
\hline
Security Metrics Maximization & \cite{Caron2013}\\
\hline
Shared Last Level Cache (SLLC) Contention Minimization & \cite{Jin2013}\\
\hline
Total Job Completion Time Minimization & \cite{li2013migration}\\
\hline
\end{tabular}
\end{table}

\subsection{Resource Utilization Maximization}
Cloud infrastructures are commonly composed by multiple physical and virtual resources such as CPU, RAM, storage, network bandwidth and Graphical Process Units (GPU). In this context, an efficient and balanced utilization of these resources is an important issue to address.
Taking into account the surveyed universe, 15.5\% of the studied articles proposed the optimization of resource utilization. Objective functions classified in this optimization group are: (1) Elasticity Maximization, (2) Maximum Average Utilization Minimization, (3) Resource Utilization Maximization and (4) Resource Wastage Minimization (see Table \ref{table_resource}).

\begin{table}[!t]
\renewcommand{\arraystretch}{1.3}
\centering
\caption{Objective Functions - Resource Utilization Maximization Group}
\label{table_resource}
\begin{tabular}{|m{5cm}|m{2.5cm}|}
\hline
\bfseries Objective Function & \bfseries References\\
\hline
Elasticity Maximization & \cite{Li2013a}\\
\hline
Maximum Average Utilization Minimization & \cite{adamuthe2013multiobjective,guo2013shadow,Li2012a}\\
\hline
Resource Utilization Maximization & \cite{song2014optimization,cao2014energy,Sun2013,hwang2013hierarchical,Jin2012,Calcavecchia2012,Mishra2011}\\
\hline
Resource Wastage Minimization & \cite{Gao2013,Xu2010,adamuthe2013multiobjective}\\
\hline
\end{tabular}
\end{table}

\begin{table}[!b]
\renewcommand{\arraystretch}{1.3}
\caption{Objective Functions}
\label{table_objective_functions}
\begin{tabular}{|m{2.3cm}|m{1.2cm}|m{4cm}|}
\hline
\bfseries Objective Function & \bfseries Percentage & \bfseries References\\
\hline
\centering Energy Consumption Minimization & \centering 50.0\% & \cite{tang2014hybrid,wang2014eqvmp,cao2014energy,zhang2014dynamic,pires2013multi,lopez2013virtual,Sun2013,Dong2013,Gao2013,fang2013power,anand2013virtual,dong2013energy,dong2013virtual,wang2013particle,singh2013reduce,fang2013vmplanner,Shigeta2013,ribas2013pbfvmc,Huang2013,chen2013intelligent,Sato2013,li2013energy,moreno2013improved,prevost2013optimal,tsai2013prevent,dalvandi2013time,Wu2012b,Wu2012,Kantarci2012,Jiang2012,Wu2012a,Huang2012a,Jin2012,Goudarzi2012,Ribas2012,Dupont2012,Ferreto2011,Ho2011,Ferreto2011a,Machida2010,Xu2010,Gupta2008}\\
\hline
\centering Network Traffic Minimization & \centering 30.9\% & \cite{wang2014eqvmp,song2014optimization,Zhang2013,pires2013multi,lopez2013virtual,Chang2013,georgiou2013exploiting,fang2013power,alicherry2013optimizing,anand2013virtual,dong2013virtual,Shigeta2013,chen2013intelligent,kakadia2013network,hong2013qoe,Wang2012,Zamanifar2012,Biran2012,Dias2012,Jiang2012a,Calcavecchia2012,Jayasinghe2011,Ferreto2011,Ferreto2011a,Piao2010,Meng2010}\\
\hline
\centering Economical Revenue Maximization & \centering 22.6\% & \cite{zhang2014dynamic,adamuthe2013multiobjective,pires2013multi,lopez2013virtual,Dong2013,shi2013provisioning,dang2013higher,Kord2013,prevost2013optimal,hong2013qoe,dalvandi2013time,Huang2012b,Huang2012,Le2011,Shi2011,mark2011evolutionary,Xu2010,Speitkamp2010,chaisiri2009optimal}\\
\hline
\centering Performance Maximization & \centering 16.7\% & \cite{Caron2013,li2013migration,lu2013qos,Shigeta2013,Jin2013,Sato2013,gupta2013hpc,Wang2012,Lin2012,Gupta2012,Calcavecchia2012,Tsakalozos2011,Bin2011,Do2011}\\
\hline
\centering Resource Utilization Maximization & \centering 15.5\% & \cite{song2014optimization,cao2014energy,adamuthe2013multiobjective,Sun2013,Gao2013,hwang2013hierarchical,Li2013a,guo2013shadow,Li2012a,Jin2012,Calcavecchia2012,Mishra2011,Xu2010}\\
\hline
\end{tabular}
\end{table}

\section{Solution Techniques}
In the considered universe, there exist different techniques proposed for solving the VMP problem. The main solution techniques include (1) deterministic algorithms, (2) heuristics, (3) meta-heuristics, and (4) approximation algorithms. The mentioned solution techniques are detailed in the following sub-sections and summarized in Table \ref{table_solution_technique}.

\subsection{Deterministic Algorithms}
Deterministic approaches are also proposed for the VMP problem, representing 19.0\% of the studied articles. Particular methods of this solution technique are presented in Table \ref{table_deterministics}. 

\begin{table}[!t]
\renewcommand{\arraystretch}{1.3}
\centering
\caption{Solution Techniques - Deterministic Algorithms}
\label{table_deterministics}
\begin{tabular}{|m{5cm}|m{2.5cm}|}
\hline
\bfseries Technique & \bfseries References\\
\hline
Binary Integer Programming (BIP) & \cite{Li2012a}\\
\hline
Constraint Programming & \cite{Bin2011,dang2013higher}\\
\hline
Convex Optimization Theory & \cite{song2014optimization}\\
\hline
Dynamic Programming & \cite{Wu2012b,Goudarzi2012}\\
\hline
Integer Linear Programming (ILP) & \cite{anand2013virtual}\\
\hline
Linear Programming (LP) & \cite{Ferreto2011a}\\
\hline
Matrix Transformation Algorithm & \cite{Sun2013}\\
\hline
Mixed Integer Linear Programming (MILP) & \cite{Kantarci2012}\\
\hline
Primal-Dual (Hungarian) Algorithm	& \cite{alicherry2013optimizing}\\
\hline
Pseudo-Boolean Optimization (PBO)	& \cite{ribas2013pbfvmc,Ribas2012}\\
\hline
Relationship Rules Framework	& \cite{Zhang2013}\\
\hline
Stochastic Integer Programming (SIP) & \cite{chaisiri2009optimal}\\
\hline
Transverse Matrix Approach	& \cite{Piao2010}\\
\hline
\end{tabular}
\end{table}

\subsection{Heuristics}
In this work, 67.8\% of the studied articles proposed heuristic-based solution techniques for the VMP problem. Particular methods of this solution technique are based on: (1) Best Fit Decreasing (BFD), (2) Based on Best Fit (BF), (3) Based on First Fit (FF), (4) Based on First Fit Decreasing (FFD), (5) Dominant Resource First (DRF), (6) First-Come, First-Served (FCFS), (7) Heaviest First (HF) and (8) Worst Fit-based (WF). Also (9) Greedy Algorithms and (10) Novel Heuristics are proposed (see Table \ref{table_heuristics}).

\begin{table}[!t]
\renewcommand{\arraystretch}{1.3}
\centering
\caption{Solution Techniques - Heuristics}
\label{table_heuristics}
\begin{tabular}{|m{5cm}|m{2.5cm}|}
\hline
\bfseries Heuristic & \bfseries References\\
\hline
Based on Best Fit Decreasing (BFD)	& \cite{wang2014eqvmp,Ferreto2011,Dong2013,zhang2014dynamic}\\
\hline
Based on Best Fit (BF) & \cite{dong2013virtual,fang2013power}\\
\hline
Based on First Fit (FF)		& \cite{Shi2011,hwang2013hierarchical,fang2013power}\\
\hline
Based on First Fit Decreasing (FFD)		& \cite{anand2013virtual,Jin2012}\\
\hline
Dominant Resource First (DRF)		& \cite{Jin2012}\\
\hline
First-Come, First-Served (FCFS)		& \cite{moreno2013improved}\\
\hline
Greedy Algorithms		& \cite{fang2013vmplanner,Li2012a,Biran2012,Wang2012,kakadia2013network,Jayasinghe2011,dong2013energy,Le2011,guo2013shadow}\\
\hline
Heaviest First (HF)		& \cite{Ho2011}\\
\hline
Worst Fit-based (WF)		& \cite{fang2013power}\\
\hline
Novel Heuristics		& \cite{li2013migration,Jiang2012,Huang2012b,Gupta2012,Huang2012,Lin2012,Biran2012,gupta2013hpc,Sato2013,Calcavecchia2012,Huang2013,Zamanifar2012,Speitkamp2010,Do2011,Dias2012,shi2013provisioning,Dupont2012,tsai2013prevent,li2013energy,Huang2012a,Li2013a,Machida2010,Jiang2012a,li2013migration,Kord2013,cao2014energy,Chang2013,hong2013qoe,singh2013reduce,Jin2013,Jiang2012,Caron2013,Meng2010,prevost2013optimal,dalvandi2013time,Gupta2008,Mishra2011,georgiou2013exploiting,lu2013qos}\\
\hline
\end{tabular}
\end{table}

\subsection{Meta-Heuristics}
From the studied articles, 15.5\% proposed the solution of the VMP problem with meta-heuristics. Particular methods of this solution technique are: (1) Ant Colony Optimization (ACO), (2) Evolutionary Algorithm (EA), (3) Genetic Algorithm (GA), (4) Memetic Algorithm (MA), (5) Neighborhood Search (NS), (6) Particle Swarm Optimization (PSO), (7) Simulated Annealing (SA), (8) Cut-and-Search (CS) and (9) Tabu Search (TS) (see Table \ref{table_metaheuristics}).

\begin{table}[!t]
\renewcommand{\arraystretch}{1.3}
\centering
\caption{Solution Techniques - Meta-Heuristics}
\label{table_metaheuristics}
\begin{tabular}{|m{5cm}|m{2.5cm}|}
\hline
\bfseries Meta-Heuristic & \bfseries References\\
\hline
Ant Colony Optimization (ACO) & \cite{Gao2013}\\
\hline
Cut-and-Search (CS) & \cite{chen2013intelligent}\\
\hline
Evolutionary Algorithm (EA) & \cite{mark2011evolutionary}\\
\hline
Genetic Algorithm (GA) & \cite{Wu2012a,adamuthe2013multiobjective,Xu2010}\\
\hline
Memetic Algorithm (MA) & \cite{tang2014hybrid,lopez2013virtual,pires2013multi}\\
\hline
Neighborhood Search (NS) & \cite{wang2013particle,Shigeta2013}\\
\hline
Particle Swarm Optimization (PSO) & \cite{Gao2013,Xu2010,adamuthe2013multiobjective}\\
\hline
Simulated Annealing (SA) & \cite{Tsakalozos2011,Wu2012}\\
\hline
Tabu Search (TS) & \cite{Ferreto2011}\\
\hline
\end{tabular}
\end{table}

\subsection{Approximation Algorithms}
Heuristics and meta-heuristics provide good quality solutions, but quality of the expected solutions is hardly measurable. In a $p$-approximation algorithm, the value of a solution will not be more (or less) than a factor $p$ times the optimum solution. Only 2.4\% of the studied articles proposed approximation algorithms for solving the VMP problem \cite{fang2013vmplanner,Wu2012b}.

\begin{table}[!t]
\renewcommand{\arraystretch}{1.3}
\caption{Solution Techniques}
\label{table_solution_technique}
\begin{tabular}{|m{2.4cm}|m{1.2cm}|m{3.9cm}|}
\hline
\bfseries Solution Technique & \bfseries Percentage & \bfseries References\\
\hline
\centering Deterministic Algorithms & \centering 19.0\% & \cite{song2014optimization,Zhang2013,Sun2013,alicherry2013optimizing,anand2013virtual,dang2013higher,ribas2013pbfvmc,Wu2012b,Kantarci2012,Li2012a,Goudarzi2012,Ribas2012,Bin2011,Ferreto2011a,Piao2010,chaisiri2009optimal}\\
\hline
\centering Heuristics & \centering 67.8\% & \cite{wang2014eqvmp,cao2014energy,zhang2014dynamic,Chang2013,Caron2013,li2013migration,georgiou2013exploiting,lu2013qos,Dong2013,fang2013power,anand2013virtual,dong2013energy,dong2013virtual,singh2013reduce,shi2013provisioning,fang2013vmplanner,hwang2013hierarchical,Huang2013,Kord2013,Jin2013,Sato2013,Li2013a,li2013energy,gupta2013hpc,guo2013shadow,moreno2013improved,kakadia2013network,prevost2013optimal,tsai2013prevent,hong2013qoe,dalvandi2013time,Wang2012,Zamanifar2012,Huang2012b,Li2012a,Jiang2012,Lin2012,Biran2012,Huang2012a,Huang2012,Dias2012,Jin2012,Jiang2012a,Gupta2012,Calcavecchia2012,Dupont2012,Jayasinghe2011,Le2011,Ferreto2011,Mishra2011,Do2011,Ho2011,Shi2011,Machida2010,Speitkamp2010,Meng2010,Gupta2008}\\
\hline
\centering Meta-Heuristics & \centering 15.5\% & \cite{tang2014hybrid,adamuthe2013multiobjective,pires2013multi,lopez2013virtual,Gao2013,wang2013particle,Shigeta2013,chen2013intelligent,Wu2012,Wu2012a,Tsakalozos2011,mark2011evolutionary,Xu2010}\\
\hline
\centering Approximation Algorithms & \centering 2.4\% & \cite{fang2013vmplanner,Wu2012b}\\
\hline
\end{tabular}
\end{table}

\section{Formulation: Offline or Online}
The VMP can be formulated as online or offline problems. According to this review, VMP formulations mostly are proposed as online problems (see Table \ref{table_formulations}), considering the dynamic nature of cloud computing workload.

For online formulations, workload prediction and forecasting techniques are proposed in \cite{zhang2014dynamic,fang2013vmplanner,tsai2013prevent,prevost2013optimal,moreno2013improved,Dong2013,Sato2013,Kord2013,Calcavecchia2012,Huang2012b,Kantarci2012,Goudarzi2012,Li2012a,Le2011,Do2011,mark2011evolutionary}.

\begin{table}[!t]
\renewcommand{\arraystretch}{1.3}
\centering
\caption{Formulation: Offline or Online}
\label{table_formulations}
\begin{tabular}{|m{2.9cm}|m{1.2cm}|m{3.4cm}|}
\hline
\bfseries Formulation Approach & \bfseries Percentage & \bfseries References\\
\hline
\centering Online & \centering 77.4\% & \cite{Piao2010,Zhang2013,Speitkamp2010,Wang2012,Zamanifar2012,Tsakalozos2011,Ferreto2011,Mishra2011,Do2011,Le2011,Ho2011,Ferreto2011a,Shi2011,Chang2013,li2013migration,georgiou2013exploiting,Huang2012b,alicherry2013optimizing,lu2013qos,Kantarci2012,Jiang2012a,Ribas2012,Wu2012b,Gupta2012,Dong2013,Calcavecchia2012,Wu2012,fang2013power,anand2013virtual,Huang2013,Kord2013,Jin2013,Li2012a,Sato2013,Li2013a,li2013energy,wang2014eqvmp,dong2013virtual,Lin2012,gupta2013hpc,moreno2013improved,song2014optimization,Biran2012,Huang2012,kakadia2013network,prevost2013optimal,Dias2012,chen2013intelligent,wang2013particle,tsai2013prevent,singh2013reduce,hong2013qoe,shi2013provisioning,Jin2012,guo2013shadow,dalvandi2013time,fang2013vmplanner,Shigeta2013,Dupont2012,Goudarzi2012,dang2013higher,cao2014energy,ribas2013pbfvmc,zhang2014dynamic,hwang2013hierarchical}\\
\hline
\centering Offline & \centering 22.6\% & \cite{Machida2010,adamuthe2013multiobjective,pires2013multi,lopez2013virtual,chaisiri2009optimal,Jayasinghe2011,Caron2013,tang2014hybrid,Bin2011,Sun2013,Gao2013,Xu2010,dong2013energy,Jiang2012,Wu2012a,Gupta2008,Huang2012a,Meng2010,mark2011evolutionary}\\
\hline
\end{tabular}
\end{table}

\section{Cloud Architectures}
According to this review, the VMP problem have been studied in 3 different cloud architectures: (1) single-cloud (i.e. only one cloud computing datacenter), (2) multi-cloud (i.e. many cloud computing datacenters from one o more cloud service providers), and (3) federated clouds. The most studied cloud architecture is the single-cloud (see Table \ref{table_architectures}).

\begin{table}[!t]
\renewcommand{\arraystretch}{1.3}
\centering
\caption{Cloud Architectures}
\label{table_architectures}
\begin{tabular}{|m{2.9cm}|m{1.2cm}|m{3.4cm}|}
\hline
\bfseries Cloud Architecture & \bfseries Percentage & \bfseries References\\
\hline
\centering Single-Cloud & \centering 88.1\% & \cite{fang2013vmplanner,Wu2012b,Li2012a,dang2013higher,song2014optimization,Goudarzi2012,anand2013virtual,Ferreto2011a,Sun2013,alicherry2013optimizing,ribas2013pbfvmc,Ribas2012,Zhang2013,Piao2010,Huang2012b,Gupta2012,Huang2012,Lin2012,Biran2012,gupta2013hpc,Sato2013,Calcavecchia2012,Huang2013,wang2014eqvmp,Ferreto2011,dong2013virtual,zhang2014dynamic,Speitkamp2010,Do2011,Dias2012,shi2013provisioning,tsai2013prevent,li2013energy,Shi2011,Jin2012,hwang2013hierarchical,fang2013power,moreno2013improved,Dong2013,Wang2012,kakadia2013network,Jayasinghe2011,dong2013energy,guo2013shadow,Ho2011,Huang2012a,Li2013a,Machida2010,Jiang2012a,li2013migration,Kord2013,cao2014energy,Chang2013,hong2013qoe,singh2013reduce,Jin2013,Jiang2012,Caron2013,Meng2010,prevost2013optimal,dalvandi2013time,Gupta2008,Mishra2011,georgiou2013exploiting,lu2013qos,Gao2013,Wu2012a,adamuthe2013multiobjective,Xu2010,tang2014hybrid,lopez2013virtual,pires2013multi,Shigeta2013,wang2013particle,Tsakalozos2011,Wu2012}\\
\hline
\centering Multi-Cloud & \centering 10.7\% & \cite{Bin2011,Kantarci2012,chaisiri2009optimal,Zamanifar2012,Le2011,mark2011evolutionary,chen2013intelligent}\\
\hline
\centering Federated & \centering 1.2\% & \cite{Dupont2012}\\
\hline
\end{tabular}
\end{table}

\section{Orientation: Provider-oriented or Broker-oriented}
Cloud infrastructure optimization is a main concern for CSPs, considering the objective functions described in Section \ref{objective_functions}. In this context, the VMP problem is commonly formulated from a provider-oriented perspective \cite{pires2013multi}.

Considering that placement decisions only include tenant’s constraints commonly represented as SLAs, tenants cannot decide which PMs will hosts their VMs. However, the number of CSPs has been rapidly increased and nowadays there are different pricing schemes, virtual machine offers and features. In general, it is difficult for users to search cloud prices and decide where to host their resources. A scenario for the VMP problem for optimizing user’s virtual infrastructure placements across available public cloud offers can be also studied. This novel scenario can be described as a broker-oriented approach \cite{chaisiri2009optimal}.

\begin{table}[!t]
\renewcommand{\arraystretch}{1.3}
\centering
\caption{Orientation: Provider or Broker}
\label{table_orientation}
\begin{tabular}{|m{2.9cm}|m{1.2cm}|m{3.4cm}|}
\hline
\bfseries Orientation & \bfseries Percentage & \bfseries References\\
\hline
\centering Provider-oriented & \centering 97.6\% & \cite{Piao2010,Zhang2013,Speitkamp2010,Wang2012,Meng2010,Zamanifar2012,Tsakalozos2011,Machida2010,adamuthe2013multiobjective,Ferreto2011,Mishra2011,Do2011,Le2011,Ho2011,Ferreto2011a,Shi2011,pires2013multi,lopez2013virtual,Chang2013,Jayasinghe2011,Caron2013,li2013migration,georgiou2013exploiting,Huang2012b,alicherry2013optimizing,lu2013qos,tang2014hybrid,Kantarci2012,Bin2011,Jiang2012a,Ribas2012,Wu2012b,Gupta2012,Sun2013,Dong2013,Calcavecchia2012,Wu2012,Gao2013,fang2013power,anand2013virtual,Xu2010,Huang2013,Kord2013, Jin2013, dong2013energy, Li2012a, Sato2013, Li2013a, li2013energy, wang2014eqvmp, dong2013virtual, Jiang2012, Wu2012a, Lin2012, gupta2013hpc, moreno2013improved, song2014optimization, Biran2012, Gupta2008, Huang2012, kakadia2013network, prevost2013optimal, Huang2012a, Dias2012, chen2013intelligent, wang2013particle, tsai2013prevent, singh2013reduce, hong2013qoe, shi2013provisioning, Jin2012, guo2013shadow, dalvandi2013time, fang2013vmplanner, Shigeta2013, Dupont2012, Goudarzi2012, dang2013higher, cao2014energy, ribas2013pbfvmc, zhang2014dynamic, hwang2013hierarchical}\\
\hline
\centering Broker-oriented & \centering 2.4\% & \cite{mark2011evolutionary,chaisiri2009optimal}\\
\hline
\end{tabular}
\end{table}

\section{Experimental Environment}
Observations about the experimental environment considered by each studied articles are presented in this section, (1) focusing on what type of experiment proposed (i.e. simulation or implementation in real applications), (2) workload type (i.e. real or synthetic), (3) workload distribution, and (4) experiment size (i.e. number of PMs and VMs).

Simulations are widely used in the studied articles, and some works even proposed a hybrid environment considering simulations and complementing
those simulations with implementations in cloud operating systems to validate the feasibility of the proposal in real scenarios \cite{Gupta2012,Jin2013,gupta2013hpc,lu2013qos}. Simulations can include experimental tests with different numbers of VMs and PMs, different types of workloads (e.g. CPU-intensive, network-intensive) and different distribution of workload (e.g. Gaussian or normal distribution). Table \ref{table_experiment_type} presents studied articles that proposed both simulations and implementations in real applications, while Tables \ref{table_workload_type} and \ref{table_workload_distribution} present workload types and distribution proposed in the studied articles. Finally, Figure \ref{figure_experiment_size} classifies articles by experiment size.

Most of the studied articles assume homogeneous resource configurations for PMs [33]. \cite{Shi2011,Biran2012,Wang2012,Jin2012,Jiang2012,Gao2013,Huang2013,Shigeta2013}.

\begin{table}[!t]
\renewcommand{\arraystretch}{1.3}
\centering
\caption{Articles considering simulations or implementation in real applications in the studied universe of 84 papers.}
\label{table_experiment_type}
\begin{tabular}{|m{2.9cm}|m{4.6cm}|}
\hline
\bfseries Experiment Type & \bfseries References\\
\hline
\centering Simulation & \cite{Gupta2008,chaisiri2009optimal,Speitkamp2010,Piao2010,Meng2010,Xu2010,Machida2010,mark2011evolutionary,Bin2011,Jayasinghe2011,Ferreto2011,Do2011,Le2011,Ho2011,Ferreto2011a,Shi2011,Tsakalozos2011,Wu2012,Biran2012,Huang2012,Wang2012,Dupont2012,Zamanifar2012,Jin2012,Jiang2012,Huang2012a,Wu2012a,Goudarzi2012,Kantarci2012,Lin2012,Jiang2012a,Ribas2012,Dias2012,Huang2012b,Li2012a,Calcavecchia2012,Wu2012b,Sun2013,Gao2013,Chang2013,Huang2013,Kord2013,Caron2013,Shigeta2013,Li2013a,li2013energy,Dong2013,dong2013energy,georgiou2013exploiting,hwang2013hierarchical,dang2013higher,gupta2013hpc,moreno2013improved,chen2013intelligent,li2013migration,adamuthe2013multiobjective,pires2013multi,kakadia2013network,prevost2013optimal,alicherry2013optimizing,wang2013particle,ribas2013pbfvmc,fang2013power,tsai2013prevent,shi2013provisioning,hong2013qoe,lu2013qos,singh2013reduce,guo2013shadow,dalvandi2013time,dong2013virtual,anand2013virtual,lopez2013virtual,fang2013vmplanner,tang2014hybrid,cao2014energy,song2014optimization,zhang2014dynamic,wang2014eqvmp}\\
\hline
\centering Applications & \cite{Gupta2012,Zhang2013,Jin2013,gupta2013hpc,lu2013qos}\\
\hline
\centering Not Identified & \cite{Mishra2011,Sato2013}\\
\hline
\end{tabular}
\end{table}

\begin{table}[!t]
\renewcommand{\arraystretch}{1.3}
\centering
\caption{Articles considering synthetic or real workload in the studied universe of 84 papers.}
\label{table_workload_type}
\begin{tabular}{|m{2.9cm}|m{4.6cm}|}
\hline
\bfseries Workload Type & \bfseries References\\
\hline
\centering Synthetic & \cite{chaisiri2009optimal,Piao2010,Machida2010,mark2011evolutionary,Bin2011,Jayasinghe2011,Ho2011,Shi2011,Tsakalozos2011,Wu2012,Biran2012,Huang2012,Wang2012,Dupont2012,Zamanifar2012,Jin2012,Jiang2012,Huang2012a,Wu2012a,Goudarzi2012,Lin2012,Dias2012,Gupta2012,Huang2012b,Li2012a,Calcavecchia2012,Wu2012b,Gao2013,Zhang2013,Caron2013,Shigeta2013,hwang2013hierarchical,dang2013higher,gupta2013hpc,chen2013intelligent,li2013migration,adamuthe2013multiobjective,pires2013multi,kakadia2013network,alicherry2013optimizing,wang2013particle,fang2013power,shi2013provisioning,hong2013qoe,lu2013qos,guo2013shadow,dalvandi2013time,dong2013virtual,anand2013virtual,lopez2013virtual,tang2014hybrid,cao2014energy,song2014optimization,zhang2014dynamic,wang2014eqvmp,tsai2013prevent}\\
\hline
\centering Real & \cite{Speitkamp2010,Xu2010,Ferreto2011,Do2011,Le2011,Ferreto2011a,Ribas2012,Sun2013,moreno2013improved,prevost2013optimal,ribas2013pbfvmc,singh2013reduce,fang2013vmplanner,tsai2013prevent}\\
\hline
\centering Not Identified & \cite{Gupta2008,Meng2010,Mishra2011,Kantarci2012,Jiang2012a,Chang2013,Huang2013,Kord2013,Jin2013,Sato2013,Li2013a,li2013energy,Dong2013,dong2013energy,georgiou2013exploiting}\\
\hline
\end{tabular}
\end{table}

\begin{table}[!t]
\renewcommand{\arraystretch}{1.3}
\centering
\caption{Articles considering different workload distributions in the studied universe of 84 papers.}
\label{table_workload_distribution}
\begin{tabular}{|m{2.9cm}|m{4.6cm}|}
\hline
\bfseries Workload Distribution & \bfseries References\\
\hline
\centering Random & \cite{Wang2012,Jin2012,Wu2012a,Huang2012b,Wu2012b,Gao2013,hwang2013hierarchical,dong2013virtual,anand2013virtual,tang2014hybrid,song2014optimization}\\
\hline
\centering Uniform & \cite{chaisiri2009optimal,Xu2010,mark2011evolutionary,Bin2011,Huang2012,Goudarzi2012,Huang2012b,Li2012a,anand2013virtual}\\
\hline
\centering Normal & \cite{chaisiri2009optimal,mark2011evolutionary,Huang2012a,Dias2012}\\
\hline
\centering Poisson & \cite{Calcavecchia2012,guo2013shadow,dalvandi2013time}\\
\hline
\centering Gaussian & \cite{Huang2012b}\\
\hline
\centering Custom & \cite{chaisiri2009optimal}\\
\hline
\centering Not Identified & \cite{Gupta2008,Speitkamp2010,Piao2010,Meng2010,Machida2010,Jayasinghe2011,Ferreto2011,Mishra2011,Do2011,Le2011,Ho2011,Ferreto2011a,Shi2011,Tsakalozos2011,Wu2012,Biran2012,Dupont2012,Zamanifar2012,Jiang2012,Kantarci2012,Lin2012,Jiang2012a,Ribas2012,Gupta2012,Sun2013,
Chang2013,Zhang2013,Huang2013,Kord2013,Jin2013,Caron2013,Shigeta2013,Sato2013,Li2013a,li2013energy,Dong2013,dong2013energy,georgiou2013exploiting,dang2013higher,gupta2013hpc,moreno2013improved,chen2013intelligent,li2013migration,adamuthe2013multiobjective,pires2013multi,kakadia2013network,prevost2013optimal,alicherry2013optimizing,wang2013particle,ribas2013pbfvmc,fang2013power,tsai2013prevent,shi2013provisioning,hong2013qoe,lu2013qos,singh2013reduce,lopez2013virtual,fang2013vmplanner,cao2014energy,zhang2014dynamic,wang2014eqvmp}\\
\hline
\end{tabular}
\end{table}

\begin{figure}[!t]
\centering
\newcommand{\slice}[4]{
  \pgfmathparse{0.5*#1+0.5*#2}
  \let\midangle\pgfmathresult

  % slice
  \draw[thick,fill=black!10] (0,0) -- (#1:1) arc (#1:#2:1) -- cycle;

  % outer label
  \node[label=\midangle:#4] at (\midangle:1) {};

  % inner label
  \pgfmathparse{min((#2-#1-10)/110*(-0.3),0)}
  \let\temp\pgfmathresult
  \pgfmathparse{max(\temp,-0.5) + 0.8}
  \let\innerpos\pgfmathresult
  \node at (\midangle:\innerpos) {#3};
}

\begin{tikzpicture}[scale=2]

\newcounter{c}
\newcounter{d}
\foreach \p/\t in {13/{1-100 VMs}, 31/{100-1000 VMs}, 19/{1000+ VMs}, 37/{Not Identified}}
  {
    \setcounter{c}{\value{d}}
    \addtocounter{d}{\p}
    \slice{\thec/100*360}
          {\thed/100*360}
          {\p\%}{\t}
  }

\end{tikzpicture}
\caption{Percentage of articles considering number of VMs used for experiments in the studied universe of 84 papers.}
\label{figure_experiment_size}
\end{figure}
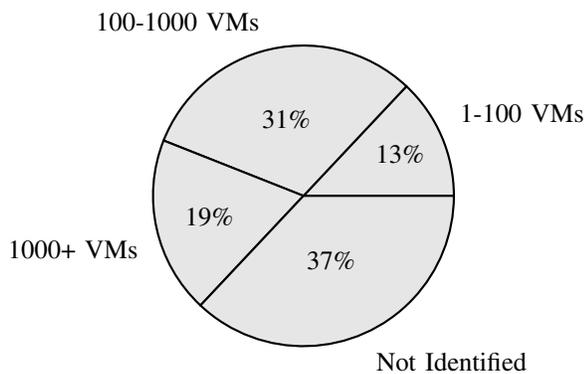

\section{Conclusions}
The presented work reviewed a large part of the VMP literature considering a well-defined selection process of research articles. Articles are classified considering: (1) optimization approaches, (2) objective functions and (3) solution techniques.

Additional identified classification criteria could be: (1) if the VMP problem is studied online or offline, (2) which cloud architectures are considered, (3) if the VMP is broker-oriented or provider-oriented and (5) what type of experimental environment is proposed.

Considering the classification and statistical data presented in this work, a deeper study of the selected articles should be performed as future direction of this PhD research.

% Close indentation from first section
}

\bibliographystyle{IEEEtranS}
\bibliography{Taxonomy}

\end{document}